\documentclass[aps,pra,reprint]{revtex4-2}
\usepackage{graphicx,bm,amsmath,amssymb}   
\usepackage[colorlinks=true, citecolor=magenta, linkcolor=blue, urlcolor=magenta]{hyperref}

\begin{document}



\title{Engineered non-Gaussian Coherence as a Thermodynamic Resource for Quantum Batteries}

\author{Kingshuk Adhikary$^{1,2}$}
\email{kingshuk.adhikary@upol.cz}
\affiliation{$^1$Department of Optics, Palack\'{y} University, 17. listopadu 1192/12, 771 46 Olomouc, Czech Republic, \\$^2$Optics and Quantum Information Group, The Institute of Mathematical Sciences,
HBNI, C. I. T. Campus, Taramani, Chennai 600113, India}

\begin{abstract}
    
Accessing quantum advantage (QA) is a legitimate task in energy harvesting devices, and it is potentially reshaping thermodynamic concepts. In this respect, the resourceful quantum non-Gaussian (QNG) states are promising candidates that precisely enable universal quantum operations to enhance thermodynamic performance with capabilities beyond what Gaussian states can achieve. We recently proposed [K. Adhikary, D. W. Moore, and R. Filip, {\em Quantum Sci. Technol.} \textbf{10}, 035048 (2025)] the QNG state generation scheme, which serves as the framework for this study and is directly integrated into the battery setting to figure out QA. By leveraging coherence in the engineered QNG states, we aim to optimize the performance of quantum batteries for various Gaussian charger profiles under unitary dynamics. We further exploit the degree of thermal broadening and environmental coupling to the charger, which is capable of fostering stable performance under precise thermal management. This study provides a proof-of-concept for exploiting thermodynamic resources in quantum energy storage units.
\end{abstract}

\maketitle

\section{Introduction}

Over the last decade, quantum advantage (QA) \cite{RevModPhys,huang2025,PhysRevResearch.5.013155,Hou2025-wy,kzvn-dj7v} has emerged as a promising frontier in the development of energy storage devices \cite{PhysRevE.87.042123, 6kwv-z6fx,Hou2025-wy,kzvn-dj7v,PhysRevResearch.5.013155,Mitchison2021,polo2025, PhysRevLett.122.047702} centered around quantum non-Gaussian (QNG) states \cite{PhysRevA.76.042327, lvovsky2020,PRXQuantum.1.020305, PRXQuantum.2.030204,LACHMAN2022100395,therm}. The remarkable features of QNG states, such as non-classical signature and Wigner negativity, provide valuable resources for advanced applications such as fault-tolerant quantum computation \cite{Bourassa2021,PRXQuantum.2.030325}, quantum error correction \cite{PhysRevX.6.031006,CAI202150}, sensing \cite{10750411,PhysRevLett.134.180801}, precision measurement \cite{PhysRevLett.128.150501,ullah2025,PRXQuantum.3.030347}, and communication and cryptography \cite{Borelli2016,Lee2019}, which are unlikely to be revived even under certain Gaussian state protocols. While the central concept is principally based in the photonic version of the QNG states of light, it has continuously been refining quantum state engineering \cite{PhysRevLett.134.180801, PhysRevLett.124.063605,crescimanna2025,PhysRevA.100.052301,PhysRevX.14.011013} to account for computational advancements \cite{PhysRevX.14.011013, PhysRevResearch.6.043212,kurman2025} that are simply unattainable with Gaussian states. In essence, QNG states in continuous variable settings \cite{kzvn-dj7v, Hou2025-wy,Mitchison2021,dowling} offer new avenues for unlocking QA, which could potentially enhance the performance of thermodynamic systems, specifically quantum batteries (QBs) \cite{6kwv-z6fx, polo2025, kurman2025}.

Several strategies \cite{lvovsky2020, PhysRevA.76.042327,PhysRevA.78.060303,PRXQuantum.2.020333,checchinato2024,PhysRevLett.134.180801,PRXQuantum.3.030347,crescimanna2025,PhysRevA.100.052301,Straka2018} have already been developed; however, they are especially useful for certifying expensive QNG states, all of which are required to unlock advanced capabilities in future tasks \cite{kurman2025,Bourassa2021,PhysRevX.6.031006,PhysRevLett.134.180801}. To avoid such engineered rigidity for the preparation of QNG states, we already proposed \cite{Adhikary_2025} a low-cost strategy that leads to a breakthrough in calibrating QNG states carrying resources as a coherence \cite{PhysRevLett.134.233604,5nxx-r97j}. This approach functions as a non-trivial QNG state generation scheme and is directly integrated into this study to offer a prototype QB setup for accessing QA \cite{6kwv-z6fx, polo2025, kurman2025}. With this development, our promising framework for the QB in this work is compelling for accessing QA without complexity, which ensures robustness to resiliently capture QNG states even with incoherent charger preparation \cite{PhysRevResearch.5.013155,6kwv-z6fx,e23050612,PhysRevResearch.5.L032010,PhysRevLett.122.047702}. By capitalizing on energy non-conservation due to minimal effort and without the requirement of an external drive, it potentially offers reliable charging mechanisms of QB under unitary dynamics. In contrast to the close QB using the conventional Jaynes–Cummings (JC) model, the energy exchange pathways from the charger to the battery create a complex dressed-state structure that likely influences the charging process. Regardless of $[\mathcal{H}_{\text {int}},\mathcal{H}_{\text {B}}+\mathcal{H}_{\text {C}}]\neq0$, total energy is conserved at the global level in a closed system with a tailored conserved operator, yet subsystem energies are not conserved separately, as with the prototype $k-$th order JC interaction \cite{vogel} under the non-zero detuning.

In this presentation, we address the implementation of nontrivial QNG states yielding QA that is capable of improving the thermodynamic performance \cite{6kwv-z6fx, Mitchison2021, RevModPhys, PhysRevResearch.5.013155,Hou2025-wy} of the QB. Without complicating it, we want to start with a Fock state charger that encodes quantum coherence \cite{Adhikary_2025,5nxx-r97j,PhysRevLett.134.233604} deterministically to facilitate key performance metrics \cite{RevModPhys,e23050612}: average charging power, extractable work, and charging efficiency. For clarity, this task enhances the overall functionality of the QB and systematically opens an avenue for optimization using the Gaussian charger \cite{6kwv-z6fx,e23050612,PhysRevResearch.5.L032010,PhysRevLett.122.047702}, which has the same mean energy. Using the conventional Gaussian-type chargers, like coherent, thermal, and squeezed states, we are able to highlight the charging performance of the battery. Our study confirms that a properly designed QNG state is optimal for the functioning of the QB, leading to maximal performance when used as a coherent resource in thermodynamic tasks \cite{6kwv-z6fx, Mitchison2021, RevModPhys,PhysRevResearch.5.013155,Hou2025-wy}. We are also leveraging thermal broadening \cite{therm} and environmental dissipation as operating resources, which enhances our understanding of non-equilibrium dynamics \cite{Mitchison2021,PhysRevLett.122.210601,PhysRevA.97.013851,dowling,PhysRevA.109.052206} and serves as a benchmark for evaluating the robustness of our proposed framework.



\section{PRELIMINARIES}
\subsection{Superposed absorption}

To start, let us remind the reader a little bit about the minimal approach \cite{Adhikary_2025} with deterministic protocol enabled for fostering the QNG coherence states \cite{PhysRevLett.134.233604,5nxx-r97j} by combining coherence-free $k-$th order JC interaction. This conditional requirement does not allow for the conservation of the number of excitations; rather, it causes an incommensurate energy exchange in the system due to frequency frustration between competing processes. This mutually active energy transformation provides an intrinsic source of bosonic coherence, leading to deterministic QNG states from initially incoherent subsystem preparations. Remarkably, in our setting, achieving the QNG state for any $k$-th JC interaction is quite feasible; however, the likelihood of maintaining coherence will be washed out. To avoid this, our effort on the superposition of the $k-$th order JC interaction is necessary in the performance of QB.

We directly employed the combination of linear ($k=1$) and nonlinear ($k=2$) absorption channels from the JC interaction, resulting in a superposed interaction Hamiltonian \cite{Adhikary_2025} of the following form (we set $\hbar=1$):
\begin{equation}
    \mathcal{H}_{\text {int}}=\sum_{k=1,2} g^{(k)}(\sigma_-a^{\dagger k}+\sigma_+a^k),
    \label{superposed}
\end{equation}
where $a$ $(a^\dagger)$ are the annihilation (creation) operators of the cavity, $\sigma_\pm$ are the ladder operators of the qubit, $g^{(k)}$ is a coupling constant. 

This Hamiltonian describes a single- and two-photon JC interaction in which the qubit (battery) exchanges simultaneously one and two photons with the oscillator (charger). It is thus meaningful to understand the possible physical manifestations of this engineered hybrid interaction in advanced setups embracing non-linearity, including circuit QED \cite{Deppe2008}, NV-center systems \cite{PhysRevLett.121.123604}, and trapped-ion architectures \cite{RevModPhys.75.281}. The last one is the most promising platform for realizing Eq. (\ref{superposed}) by driving the first and second motional sidebands of the ion in a controlled manner. Researchers have already separately studied \cite{6kwv-z6fx,e23050612,e25030430,PhysRevResearch.6.023136,PhysRevA.97.013851,Zahia31122025} remarkable findings on thermodynamic performance using either the one- or two-photon JC model; to the best of our knowledge, their combination has yet to be addressed. This study mitigates this gap by providing a comprehensive study with minimal effort—superposed JC interactions (\ref{superposed}) that are meaningful for the desired figures of merit in the QB.


\subsection{Theoretical framework}

In the hybrid framework that includes both the oscillator and the qubit, the oscillator acts as a `charger' (C) while the qubit serves as a `battery,' (B) as defined by their respective Hamiltonians below:
\begin{eqnarray}
    \mathcal{H}_\text{C}&=&\omega_ca^\dagger a,\\
    \mathcal{H}_\text{B}&=&\frac{\omega_q}{2}(\mathbb{I}+\sigma_z).
\end{eqnarray}
The frequency of free motions in lab-frame accounting is represented by $\omega_{c(q)}$. As a scaling unit for figures of merit in this study, we set the ground state of the qubit to zero energy and the excited state to maximum charging energy, represented by $\omega_q$. We take the initial energy of the charger in accordance with its preparation, taking into account the Gaussian profile and QNG states for a fair comparison of QA with the baseline of bosonic coherence.  We set the initial state at time $\tau = 0$ for the joint system as $\rho(0)\equiv\rho_{\text{B}}(0)\otimes\rho_{\text{C}}(0) $, where $\rho_{\text{B}}(0)=|g\rangle\langle g|$ is the zero-energy ground state of $\mathcal{H}_\text{B}$, and while $\rho_{\text{C}}(0)$ characterizes the charger $\mathcal{H}_\text{C}$ as having an initial energy determined from the charger configuration. It is important to note that the battery and charger interact only when $\tau>0$, and we assume that the superposed interaction (\ref{superposed}) is activated at this stage. This unitary dynamic enables the overall functionality and optimal thermodynamic performance of QB for a charging time interval of $0\le\tau\le2\pi$.


To investigate the QNG‑based advantage of the QB, let us consider the following key figure of merits: The energy $E(\tau)$ stored in the QB at time $\tau$ and its corresponding average charging power $\bar P$ are given by
\begin{eqnarray}
    E(\tau)&=&\text{Tr}[\mathcal{H}_\text{B}\rho_{\text{B}}(\tau)]    \label{energy}\\
    \bar P&=&\frac{\text{max}(E)}{\tau}\Bigg|_{\tau=\tau_c},
\end{eqnarray}
where $\tau_c$ is the fastest time at which the energy of QB reaches its maximum value $E_{\max}={\max}(E)$. In order to have a complete characterization of the QB, we now focus on ergotropy $\xi$, which reflects the conversion into maximum extractable work from the stored energy via unitary operations, and to better quantify the actual fraction $\eta$ of extractable energy, both are given by
\begin{eqnarray}
\xi(\tau) &=& E(\tau)-\frac{\omega_q}{2}\left[1-\sqrt{1-4\text{det}\rho_\text{B}(\tau)}\right]\\
\eta(\tau)&=&\frac{\xi(\tau)}{E(\tau)}.
\label{efficiency}
\end{eqnarray}

Understanding these figures of merit allows us to optimize strategies for harnessing the QNG advantage of QB according to charger configuration. It is worth mentioning that signal-to-noise ratio (SNR) is another quantitative indicator in this study, especially when studying charging precision and fluctuations in the stored energy $E$, which is defined as
\begin{equation}
    \text{SNR}(\tau)=\text{log}_{10}\frac{E^2(\tau)}{\Delta E^2(\tau)}.
    \label{snr}
\end{equation}
We are now ready to focus on our findings in accordance with Eqs. (\ref{energy})–(\ref{efficiency}) of this proposed theoretical framework. Notably, all system parameters are scaled by the maximum energy $\omega_q$, unless otherwise specified.

\section{RESULTS}

\subsection{Performance optimization of the QB from the Fock-based charger}

\begin{figure}[t]
\centering
  \begin{tabular}{@{}ccc@{}}
    \includegraphics[width=\columnwidth]{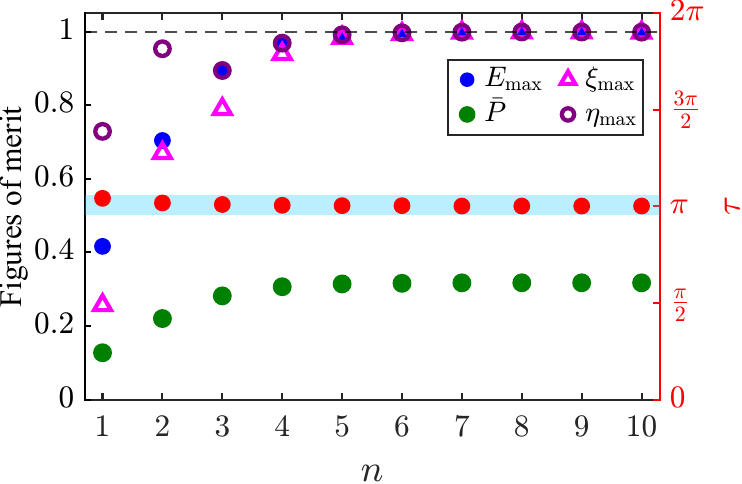} 
     \end{tabular} 
 \caption{ (Color online). Achieve a Fock state-based QA using balanced couplings ($g^{(1)}=g^{(2)}=1$) governed by a superposed charging protocol. The red points represent the measured optimal charging time $\tau_c$ at which the battery energy reaches its maximum value $E_{\max}$, as do the rest of the figures of merit that achieved their maximum. It generally occurs around $\tau=\pi$, although existing fluctuations (color region) of order $\approx 10^{-1}$ occur for some initial excitation $n\le6$. Notice that as non-classicality increases in the charger, QA reaches a saturation at $n=6$ across the range $0\le\tau\le2\pi$. The horizontal (dashed-dotted black) line represents the ideal charging condition, which is of $E_{\text{max}}=\omega_q$.}
 \label{main1}
\end{figure}

The primary task is to characterize the performance of the QB using the superposed charging protocol according to Eq. (\ref{superposed}), after preparing the charger in a generic Fock state $|n\rangle$. In principle, this straightforward approach is well-controlled, which determines the optimal configuration under which the QB can achieve a Fock-state-based advantage. The simulations indicate that the interplay of the two different coupling strengths in the balanced coupling provides optimal charge-transfer channels, as demonstrated in Figure \ref{main1}. The non-trivial resonant conditions lead to non-equilibrium charging dynamics regardless of pairs $(g^{(1)}, g^{(2)})$, but illustrate the optimizing charging protocols to enhance battery performance only when $g^{(1)}=g^{(2)}$. Furthermore, two different photon channels exchange photons between the charger and the qubit simultaneously, resulting in two channels that interfere constructively over the interval $0\le\tau\le2\pi$ when $g^{(1)}=g^{(2)}$, as opposed to $g^{(1)}\neq g^{(2)}$.  Our observations, as presented in the Supplemental Material \cite{sup}, confirm this assumption. It is worth pointing out that the QNG coherence is attains maximum around $\tau=\pi$ likely for any initial excitation $|n\rangle$ indicates the maximum energy stored into battery governed by the charging mechanism Eq. (\ref{superposed}). Note that (see Fig. \ref{main1}), the Fock states show well-charging performances showing saturation plateau beyond $n=6$ in terms of relevant figures of merit mentioned in Eqs. (\ref{energy})-(\ref{efficiency}) and reach the maximum QA with increasing nonclassicality $n$. A key point is that the emergence of maximum quantum coherence in the charger at $\tau_c$ conclusively boosts the thermodynamic performance of QB, making a Fock-based charger the ideal setup to optimize QA. From $n=7$, battery charges precisely induce maximum QA due to saturation behavior (maximum) of coherence, as discussed in Ref. \cite{Adhikary_2025}. By leveraging these configurations for Gaussian charger preparation, we can measure a definite QA like the maximum performance of existing scalable Fock-state charging configurations. To do this, we adjust the mean number of quanta $\langle n\rangle=7$ for the initial Gaussian profiles, including coherent, thermal, and squeezed states, all having optimized charging time $\tau_c=\pi$.

\begin{figure*}[t]
\centering
  \begin{tabular}{@{}ccc@{}}
    \hspace{-.3in}
    \includegraphics[scale=0.7]{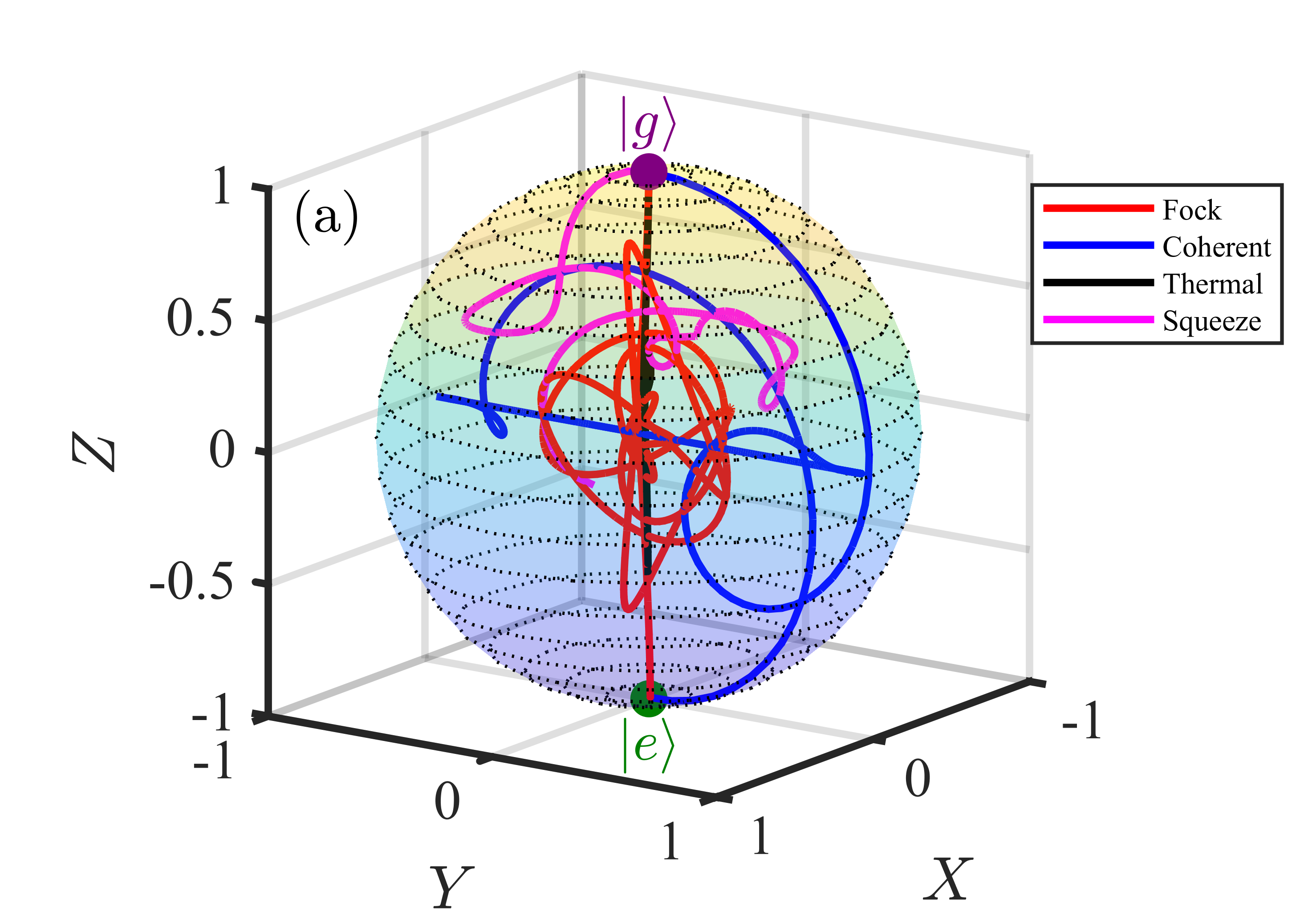} & \hspace{.1in}
    \includegraphics[scale=0.7]{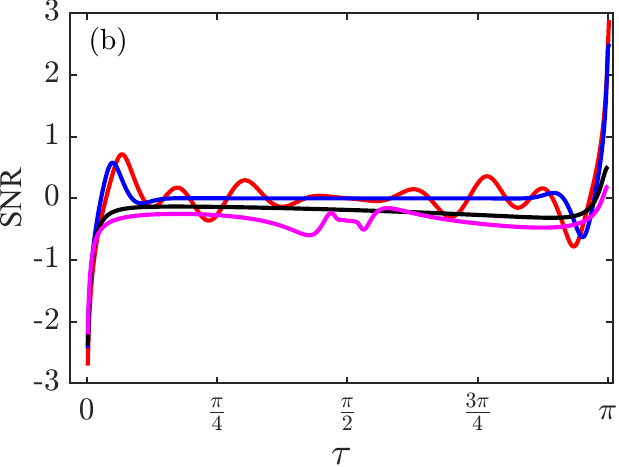}
     \end{tabular}  
 \caption{ (Color online). (a) Bloch sphere representation of the state of qubit battery evolution under different charger profiles, each with mean energy $\langle n \rangle=7$ to ensure equal average energy input to the qubit battery.  The north and south poles correspond to the empty ($|g\rangle$) and fully charged ($|e\rangle$) states, respectively. The trajectory exhibits the reduced qubit state evolving up to charging time $\tau_c$ while interacting with four distinct bosonic charger states: Fock state (red), coherent state (blue), thermal state (black), and squeezed vacuum state (magenta).  The trajectory is periodic yet non-circular because the superposed charging interaction enables energy to be exchanged simultaneously in one- and two-photon channels, resulting in frequency frustration during the charging process. During charging processes, the state of the battery evolves and traces a trajectory on or inside the Bloch sphere, reflecting changes in energy, coherence, and purity. Apart from the Fock charger, only the coherent charger is capable of completing the path between $|g\rangle$ and $|e\rangle$. (b) The signal-to-noise ratio (SNR) of the battery changes over time while it is being charged with similar chargers shown in the left panel. Despite all chargers providing the same mean energy, the SNR with higher values indicates a more reliable and less fluctuating energy-storing battery.}
 \label{bloch}
\end{figure*}

\subsection{Performances with Gaussian chargers}

Given that we are considering the diversification of chargers with coherent, thermal, and squeezed states, it is interesting to see which Gaussian charger is well-suited for maximum QA. During charger preparation, we initialize the battery in its ground state and configure the charger so that Gaussian states deliver the precise amount of energy required to charge it to its excited state. Following the key figures of merit presented in Eqs. (\ref{energy})-(\ref{efficiency}), we implement advanced simulations, such as the Bloch sphere and SNR, to visualize the optimal configurations of QB up to maximum charging. Both representations are well-suited to analyze how the various Gaussian states interact and affect performance on QB. 

Figure \ref{bloch}(a) demonstrates a visualization of the charging strategies of the battery up to the full charging cycle $\tau_c$ by leveraging the unique properties of Gaussian states. Here, the Fock charger is referenced only to understand insights into energy transfer dynamics followed from the Gaussian chargers. As expected \cite{e23050612}, the Fock charger deterministically completes a full charging cycle, indicated by the Bloch vector traveling entirely along a path from $|g\rangle$ to $|e\rangle$ of the sphere. This deterministic behavior is predictable because of the emergence of quantum coherence in the Fock charger, which can charge the qubit precisely despite energy fluctuations caused by the superposed charging protocol. This predictability does not contrast sharply with the coherent charger and allows for the perfect transfer of energy after charging due to substantial coherence capable of overcoming the unavoidable energy fluctuations. Although other Gaussian chargers cannot provide the advantage of coherence, this indicates that the battery gains insufficient energy during the superposed interaction, resulting in an incomplete Bloch trajectory between $|g\rangle$ and $|e\rangle$. The results with the existing non-trivial charging strategy emphasize the importance of coherence for an optimal charger configuration, even with uncontrolled energy transfer dynamics between the charger and the battery. This analysis identifies the most promising charger in Gaussian profiles as the coherent state, which provides a complete charging cycle, like the Fock charger. It is worth noting that the emergence of coherence in accordance with charger configurations resulting from the superposed JC interaction leads to complex Bloch trajectories along the route  $|g\rangle\rightarrow|e\rangle$. Further details of the Bloch vector are provided in the Supplemental Material \cite{sup}.

To identify the good QB with the aforementioned charger configurations, we now discuss how SNR quantifies the reliability of energy storage relative to its quantum fluctuations. Understanding SNR as defined in Eq. (\ref{snr}) in this context allows us to assess QA in the presence of fluctuation, and we can determine a QB design that delivers energy sharply. Results are displayed in Fig. \ref{bloch}(b). We observe that, in general, charging dynamics are subject to characterization of charging configurations with the degree of SNR (see figure \ref{bloch}(b)), which confirms an energetic QB is not solely determined by average energy. Regardless of charger configuration, intermediate charging provides SNR $\le0$, indicating that fluctuations are comparable and dominate over average stored energy. This insight improves as SNR $>1$ for the Fock and coherent chargers after complete charging, indicating QA with deterministic stored energy. On the contrary, thermal and squeeze chargers are both unsuitable for designing the battery under the superposed charging protocol because the coherence in the charger is insufficient to overcome SNR $=1$. This limitation emphasizes that the performance of thermal and squeeze chargers does not provide efficient energy storage solutions, likely what we obtained from Fock and thermal charger configurations.

\begin{figure}[h]
\centering
  \begin{tabular}{@{}ccc@{}}
    \includegraphics[width=\columnwidth]{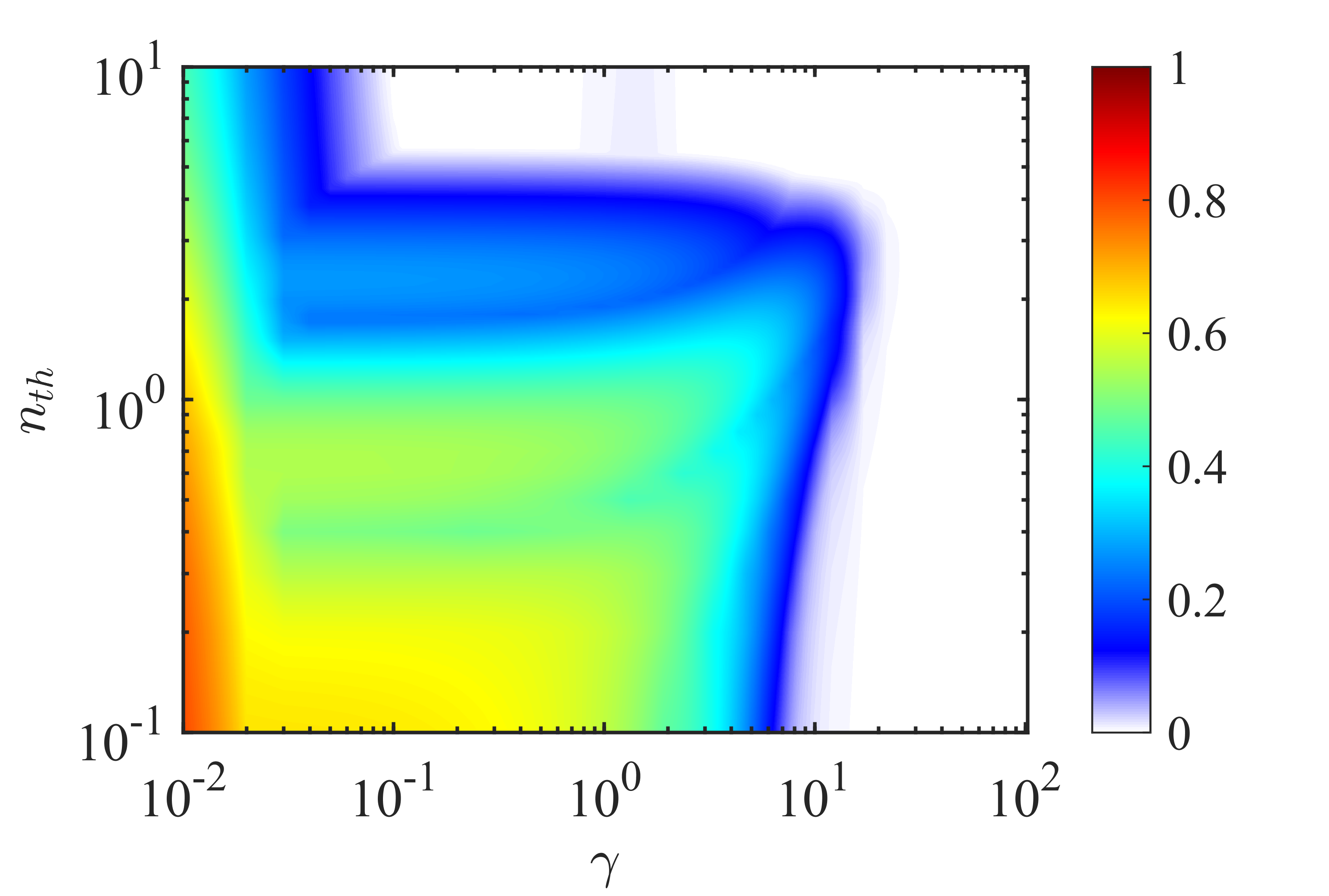} 
     \end{tabular} 
 \caption{ (Color online). The maximum ergotropy is measured over the time interval $0\le\tau\le2\pi$, which results in stable performance of the battery, indicated by the colored region. The charger is initially prepared in a thermalized Fock state ($|7\rangle$) and coupled to a dissipative environment. For large dissipation, specifically when $\gamma > 10$, the charger cannot supply energy to the battery, resulting in no additional work being extracted, as illustrated by the white region. On the contrary, the thermal spread $n_{th}>1$ does not stabilize the charger for accounting for the finite dissipation $\gamma\le0.1$. Battery performance is limited due to the transition from finite to zero ergotropy as dissipation increases. This highlights the fragility of work extraction under realistic dissipative conditions, even when the charger is prepared in a pure Fock state, $n_{th}=0$, as confirmed in the Supplemental Material \cite{sup}.}
 \label{dissipation}
\end{figure}

\subsection{Thermalized Fock charger with environmental coupling}

So far, we have calibrated the performance of the QB by employing an ideal charger configuration under superposed charging interactions during unitary evolution. Let us now consider preparing a realistic charger model, which is itself subject to environmental dissipation and has a thermal distribution at the finite temperature $n_{th}$ centered around a specific Fock state $|n\rangle$ \cite{therm}. This open-system energy source with the thermal broadening directly interacts with the battery, significantly affecting its charging performance under non-unitary evolution. This thermal-dissipative charger framework is potentially suitable to investigate battery stability by accounting for the degradation of QAs and the fundamental limits of the energy transfer mechanism under the superposed interaction. More details about the characteristics of the thermalized Fock charger are presented in the Supplemental Material \cite{sup}. It is then interesting to quantify the ergotropy \cite{PhysRevLett.122.210601,Hadipour2024,Hadipour2025,dowling,PhysRevA.109.052206} to simulate how much instantaneous work could be extracted while isolating the battery; however, this does not imply actually extracting that work while dissipation continues.

Figure \ref{dissipation} exhibits the stable performance of the QB by measuring ergotropy while the charger experiences different levels of tolerance due to thermal broadening $n_{th}$ and environmental coupling $\gamma$. Notably, these tolerances enable us to optimize the charger for accessing QA, which focused only on stable performance instead of enhancing them. The QA refers to the amount of instantaneous work extracted from the battery, accounting for the thermalized Fock charger compared to the pure Fock charger; both are subject to the environmental coupling. On the other end, the degree of stable performance is quantified as $\xi_\text{max}\ne0$ within a certain range of $\gamma$ because the thermal broadening $n_{th}\ge10$ accelerates the saturation of work extraction; see the Supplemental Material for details \cite{sup}. Consequently, thermal broadening $10^{-1}\le n_{th}\le1$ optimizes charger loss within the range of $0.01\le\gamma\le10$, indicating that some energy is transferred to the battery while the remainder is lost to the environment. By leveraging the interplay between thermalization and environmental factors, the preparation of the Fock charger becomes more robust, allowing access to QA in the battery with its stable performance. The effects of thermal broadening and environmental coupling on the Fock charger are discussed separately in the supplementary material \cite{sup} to better understand their roles in the charging process.

\section{Conclusions}
It is fair to say that the formalism for the superposed charging interaction does not represent a competition between multiphoton JC interactions and perturbative treatment. Indeed, it emphasizes the equal contribution of the $k$-th order JC interaction, which serves as a comprehensive framework for understanding the underlying processes during charging dynamics. The higher-order coherence \cite{PhysRevLett.134.233604, 5nxx-r97j} is primarily due to the nonlinear JC interaction, which is in good agreement for creating a perfect charger in a closed setting \cite{6kwv-z6fx,e23050612,polo2025}. These findings reveal a diversification of charger profiles in theoretical contexts, potentially allowing for genuine QA \cite{kzvn-dj7v,RevModPhys,PhysRevResearch.5.013155} by leveraging QNG states with substantial coherence. On the practical side, our study focuses on a tailored framework for a sustainable charger with environmental coupling, while QA is unlikely to be noticeable in the battery. Quite remarkably, the thermalized Fock state is a potential solution to the optimal design of the charger, ensuring that the performance of the battery remains consistent despite the environmental coupling.

\section{Outlook}

This study raises an open question: the requirement of a more nuanced approach regarding the possibility of finding an energy conservation operator. As a continuation, this proposal could shed light on the scalability of the charger that will lead to more efficient energy-storing units. This would require quantum state engineering \cite{PhysRevLett.134.180801, PhysRevLett.124.063605,crescimanna2025,PhysRevA.100.052301,PhysRevX.14.011013} to prepare the resourceful states \cite{Adhikary_2025,PRXQuantum.1.020305,lvovsky2020,crescimanna2025} for the charger with noise optimization. It would also be interesting to develop a strategy for extracting useful work after refining this minimal framework into a continuous variable setting \cite{kzvn-dj7v, Hou2025-wy,Mitchison2021, PhysRevA.109.052206,dowling} with environmental coupling.


\section{Acknowledgments}

The author thanks Sibasish Ghosh for fruitful discussions and insightful comments, and Pradip Laha for helpful comments.

All numerical work was done with the Python toolbox QuTiP \cite{JOHANSSON20131234}.

\section{DATA AVAILABILITY}

The data that support the findings of this study are openly available in this repository at https://doi.org/10.5281/zenodo.19509834 \cite{zen}. The repository contains all numerical datasets and code needed to reproduce the results without restriction.

\bibliographystyle{apsrev4-2}
\bibliography{ref}  

\end{document}


\beginsupplement

\title{Supplemental Material for\\Engineered non-Gaussian Coherence as a Thermodynamic Resource for Quantum Batteries}

\author{Kingshuk Adhikary$^{1,2}$}
\email{kingshuk.adhikary@upol.cz}
\affiliation{$^1$Department of Optics, Palack\'{y} University, 17. listopadu 1192/12, 771 46 Olomouc, Czech Republic, \\$^2$Optics and Quantum Information Group, The Institute of Mathematical Sciences,
HBNI, C. I. T. Campus, Taramani, Chennai 600113, India}

\maketitle

\section{Competing Rabi frequencies for the supeposed JC interaction}

Start from the lab-frame energies with the initial state $|g,n\rangle$ projected onto the two main direct ladders separately: (i) the one-photon channel leads to $|g,n\rangle \leftrightarrow |e,n-1\rangle$ with coupling $g^{(1)}\sqrt{n}$; (ii) the two-photon channel leads to $|g,n\rangle \leftrightarrow |e,n-2\rangle$ with coupling $g^{(2)}\sqrt{n(n-1)}$. These transitions reveal insights into the energy transfer mechanism that sets the natural oscillation frequencies of the respective channels.

To generalize Rabi frequencies $\Omega_{1(2)}$ for two combined channels, we have two detunings, $\Delta_{1(2)}=\omega_q-1(2)\omega_c$, which is valid for an effective two-level system. Explicitly, $\Omega_{1}=\sqrt{\Delta^2_1+(2g^{(1)})^2n}$ and $\Omega_{2}=\sqrt{\Delta^2_2+(2g^{(2)})^2n(n-1)}$, which are capable of producing beats during charging dynamics. The competing Rabi frequency for superposed JC interaction is approximately $\delta\Omega\approx|\Omega_{1}-\Omega_{2}|$. On the basis of it, it is natural to assume that the beat period is $T_{\delta\Omega}=2\pi/\delta\Omega$.

For a better understanding, we assume that $\Delta_{1(2)}\rightarrow 0$, and balanced couplings $g^{(1)}=g^{(2)}=g$. These criteria provide $\delta\Omega=2g\left|\sqrt{n+\Delta_1^2/4g^2}-\sqrt{n(n-1)+\Delta_2^2/4g^2}\right|$. The value of $T_{\delta\Omega}$ can be approximated as $T_{\delta\Omega}\approx\frac{\pi}{g}$ for moderate $n$. This suggests that the figures of the merit plateau presented in Fig. 1 in the main part of the manuscript have a reliable charging time of $\tau_c$, which possesses a maximum relative error of $4\%$.

It is noted that balanced couplings $g^{(1)}=g^{(2)}$ lead to a common sharing of pathways, allowing perfect energy transfer between the charger and battery. In contrast to imbalanced couplings $g^{(1)}\neq g^{(2)}$, no resonant hybrid ladder survives, leading to rapid oscillations depending on excitation $n$. This physical effect implies the qubit is unable to accumulate energy efficiently, and each channel tries to charge and discharge it on different timescales, with modulation features. This observation suggests that competition between one-photon and two-photon processes is not suitable to optimize the performance of the battery; otherwise, the superposed charging protocol \cite{Adhikary_2025} is likely to be treated as perturbative by either multi-photon component.

\section{charging dynamics of the battery}

\begin{figure}[t]
\centering    \includegraphics[width=\columnwidth]{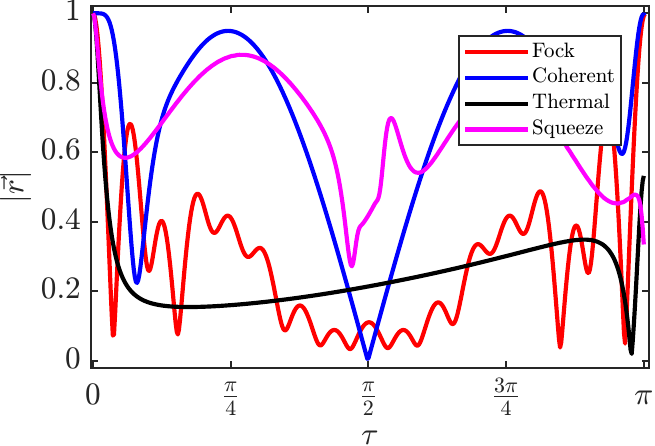} 
 \caption{ (Color online).  Time evolution of the qubit Bloch vector $|\vec{r}|$ for different initial charger states. The dynamics show the probability of finding the qubit state on or inside the Bloch sphere during charging ($0\le\tau\le\tau_c$) for a specific charger having the mean occupation number $\langle n\rangle=7$. The differences in trajectory shape describe the energy fluctuations of the charger state, which fundamentally alter the charging dynamics and coherence of the qubit battery due to the lack of a steady Rabi oscillation caused by superposed charging. Note, visualization of some complex `inside the sphere' motion of the qubit state as $|\vec{r}|<1$ reflects entanglement with the charger. }
 \label{r}
\end{figure}

The trajectory of a Bloch vector visualizes the state evolution of the qubit during charging, illustrating \cite{e23050612} how energy and coherence are transferred and stored in a battery. We optimized the potential charger that accesses QA by analyzing the paths on the Bloch sphere for the various charger configurations discussed in the main text. To extend this result, we analyze the length of the Bloch vector $|\vec{r}|$ that measures the purity and entanglement of the qubit state under a non-trivial energy charging scheme. In Figure \ref{r}, we report the behavior of $|\vec{r}(\tau)|$ for the four considered initial states of the charger. Charging processes indicate the trajectory of the Bloch vector as it moves between the north and south poles, which exhibits irregularity because the qubit does not undergo well-defined Rabi oscillations, which are affected by uncontrolled energy exchanges governed by superposed interactions. This behavior directly reflects entanglement and purity, caused by the unavoidable energy fluctuation in the different charger states, implying the resulting qubit state is no longer pure. The intermediate charging dynamics, defined for $0<\tau<\tau_c$, illustrate the inward motion of the Bloch trajectory, which represents the entanglement of the combined system and provides insight into energy swapping. As the dynamics evolve, the charger becomes mixed rapidly with the battery, which causes the energy transfer to the qubit to not be deterministic due to the inherent energy fluctuations in the charger that follow from the superposed charging interaction. After charging with the Fock and coherent state chargers, substantial coherence allows for precise battery charging, indicating the pure state of the qubit with $|\vec{r}(\tau_c)|=1$. These advancements offer to configure a potential charger for accessing QA with the interplay of coherence and battery dynamics, which is unlikely with thermal and squeezing chargers.

\begin{figure}[b]
\centering
  \begin{tabular}{@{}ccc@{}}
    \includegraphics[width=\columnwidth]{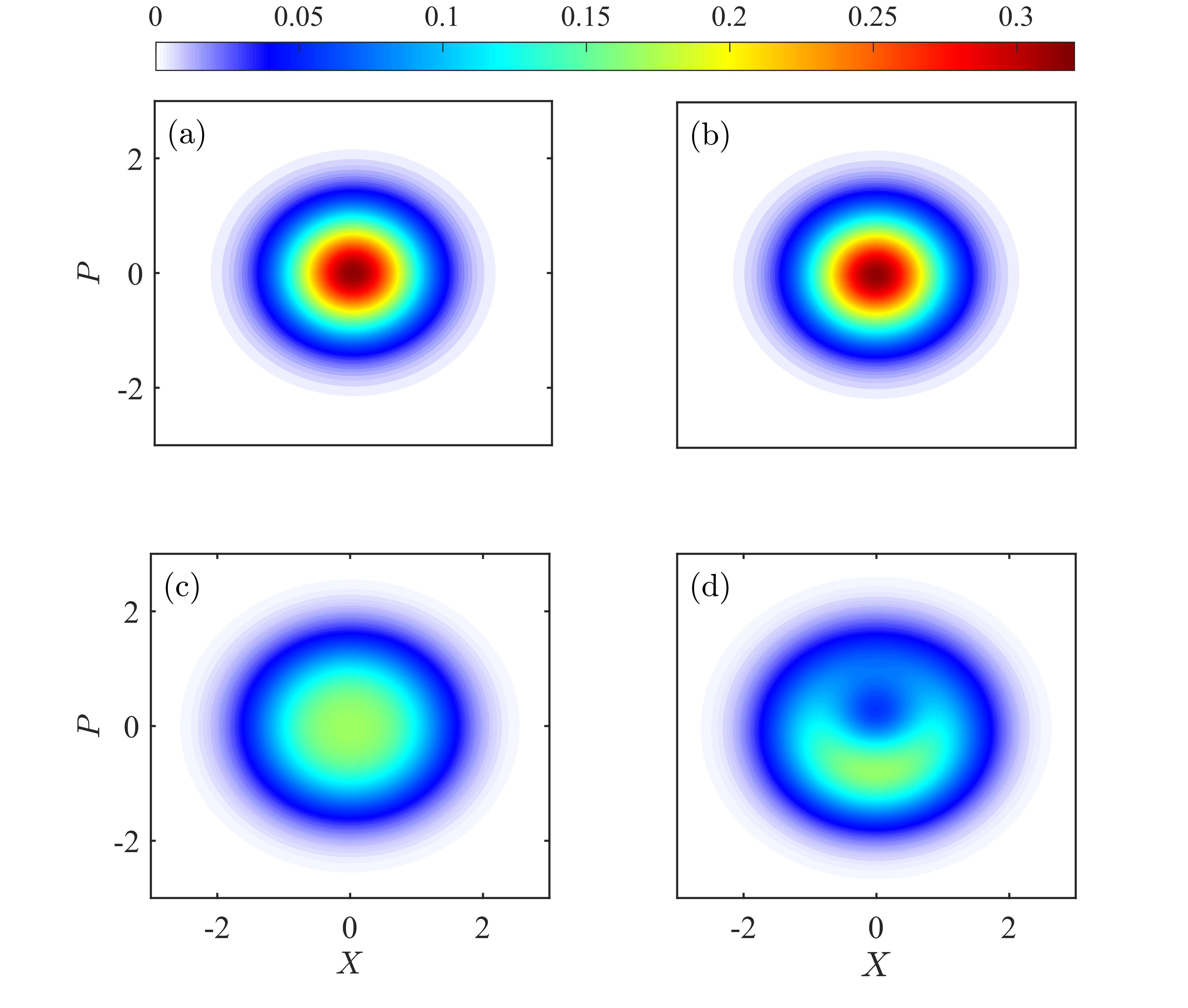} 
     \end{tabular} 
 \caption{ (Color online). Each panel displays the Wigner distribution of the reduced state of the qubit at the end of the charging period $\tau_c=\pi$, corresponding to (a) Fock state, (b) coherent state, (c) thermal state, and (d) squeezed state chargers, all with the same mean occupation number $\langle n\rangle=7$. Depending on $|\vec{r}(\tau_c)|$, the center of the Wigner function is localized in phase space, measuring the purity of the qubit state. This localization after the complete charging reveals which chargers precisely transfer energy to the battery, allowing it to reach the excited state.}
 \label{wig}
\end{figure}

Understanding and visualizing the final qubit state is essential for optimizing charger states to provide QA. Fig. \ref{wig} shows how much quantum coherence remains in the battery at $\tau_c$ for different charger initial states (Fock, coherent, thermal, squeezed). After charging with Fock and coherent chargers, the symmetric localized Wigner function justifies that the qubit is in a pure state ($|\vec{r}(\tau_c)|=1$). On the other end, thermal and squeeze chargers fail to charge the qubit; hence, the corresponding Wigner functions are no more localized, indicating the qubit is in a mixed state ($|\vec{r}(\tau_c)|<1$). This result directly evaluates the effectiveness of the entanglement between the battery and charger, which degrades the performance of the battery.


\section{Battery Performance with the Fock-State Charger under Dissipation and Thermal Noise}

\subsection{Thermalized map}

A thermalized Fock charger is represented by a statistical mixture that peaks around a specific Fock state $|n\rangle$ and exhibits thermal fluctuations. Unlike a pure Fock state, a thermalized Fock state $\rho_{|n\rangle}(n_{th})$ is influenced by thermal effects rather than being quantized. For simplicity, thermal distribution of the ground state of a harmonic oscillator better describes $\rho_{|n\rangle}(n_{th})$ \cite{therm} for its higher energy occupancies at finite temperatures $n_{th}$ and incorporates thermal uncertainties. The corresponding density matrix is a Gaussian mixture over displacements of a Fock state $|n\rangle$, indicating its thermal distribution in phase space, such that 
\begin{equation}
    \rho_{|n\rangle}(n_{th})=\frac{1}{\pi n_{th}}\int d^2\alpha e^{-\frac{|\alpha|^2}{n_{th}}}\mathcal{D}(\alpha)|n\rangle\langle n|\mathcal{D}^\dagger(\alpha).
    \label{map}
\end{equation}

We finally arrive at compact expressions that define the diagonal elements of the mixed state created by thermalizing a pure Fock state $|n\rangle$, as follows:
\begin{equation}
    \langle k|\rho_{|n\rangle}(n_{th})|k\rangle=\sum_{j=0}^{\text{min}(k,n)}\binom nj \binom kj j!\frac{n^{k+n-2j}_{th}}{(1+n_{th})^{k+n+1}},
\end{equation}
with $\sum_{k=0}^\infty \langle k|\rho_{|n\rangle}(n_{th})|k\rangle=1$. This state functions as a charger that exploits energy transfer to the battery under the superposed charging interaction. In particular, a trapped ion \cite{RevModPhys.75.281} system is a promising platform for preparing the thermalized Fock state using advanced techniques such as ground state cooling, Fock state engineering, and controlled heating. In this context, coupling to a thermal environment induces thermalization, which follows the degree of thermal broadening. It is then interesting to ask whether the degree of thermalization in the charger influences the charging dynamics because the thermal noise directly leads to fluctuations in the energy transfer process. These dynamics through the ergotropy measurement \cite{PhysRevLett.122.210601,Hadipour2024,dowling,PhysRevA.109.052206}  offer insights into optimizing both the overall functionality and the stability of the stored energy in the battery.

\begin{figure}[h]
\centering
   \includegraphics[width=\columnwidth]{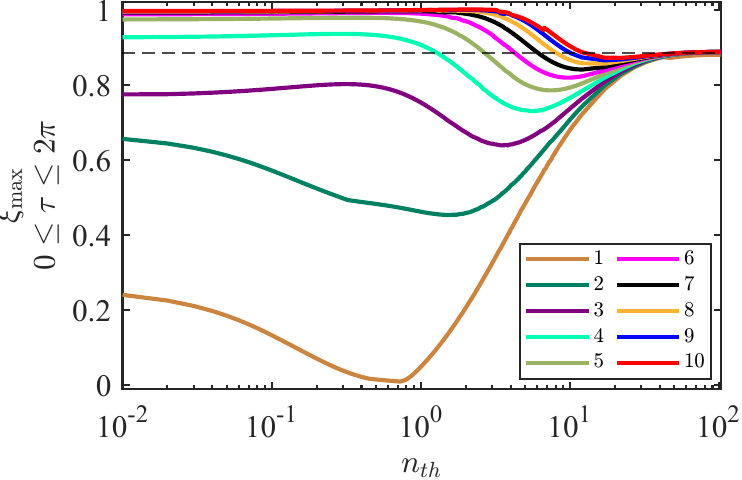} 
 \caption{ (Color online). The maximum ergotropy (in units of $\omega_q$) of the battery is measured over the charging interval $0 \le \tau \le 2\pi$, as a function of the thermal broadening $n_{th}$ (log scale) of the Fock charger. Each curve corresponds to a different initial Fock number of the charger ($n = 1$ to $10$). At very low temperatures ($n_{{th}} \ll 1$), the maximum ergotropy is determined by the initial excitation of the charger, reflecting more extractable work from higher excitations.  An increment of thermal broadening leads to a pronounced suppression of ergotropy, particularly for lower Fock numbers. The maximum ergotropy saturates around $n_{th}> 10$, indicating stable performance (black dashed horizontal line, $\xi_\text{max}=0.89\omega_q$) of the battery for all initial thermalized Fock chargers. This saturation indicates the advantages of a sustainable charger regardless of the initial state preparation.}
 \label{sup3}
\end{figure}

In Figure \ref{sup3}, we present the maximum amount of work that can be extracted through thermal broadening for various initial configurations of thermalized Fock chargers. By using the thermalized map (Eq. \ref{map}) of the Fock charger, the QB indirectly achieves stable performance due to controlled temperature broadening provided by the charger. When the thermalized Fock charger is initially prepared in the higher excitation ($n\le 6$), the battery provides perfectly maximum work extraction but saturates at a lower value with the thermal broadening. On the other end, charger preparation at lower excitation levels means the possible work extraction is not maximum but approaches the same saturation value as temperature increases. These results are suitable for realistic and sustainable charger preparation; even a lower excitation is capable of maximizing battery work at higher temperatures. In essence, the thermal spread of a pure Fock state resulting from partial quantization during unitary charging dynamics makes it a sustainable charger.

\subsection{Environmental coupling}

The ideal Fock charger with higher excitation in a closed system always enables deterministic QA \cite{e23050612,kzvn-dj7v} under the superposed charging interaction. This characteristic is presented in Fig. 1 in the main part of the manuscript. However, when the charger is inherently coupled to its surroundings, the battery is not completely charged because some of the energy leaks into the environment via the dissipation channel. The following master equation models the required non-unitary charging dynamics \cite{dowling,PhysRevLett.122.210601}:
\begin{equation}
\frac{d{\rho}}{d t} = -i[ \mathcal{H}_\text{int}, \rho] + \gamma \mathcal{D}[ a ]\rho,
\end{equation}
where $\mathcal{H}_\text{int}$ is the superposed JC interaction and $\rho$ is the state of the joint system. The Lindblad dissipator $\mathcal{D}[a]\rho$ accounts for the dissipation of the charger with decay rate $\gamma$.

\begin{figure}[h]
\centering
  \begin{tabular}{@{}ccc@{}}
    \includegraphics[width=\columnwidth]{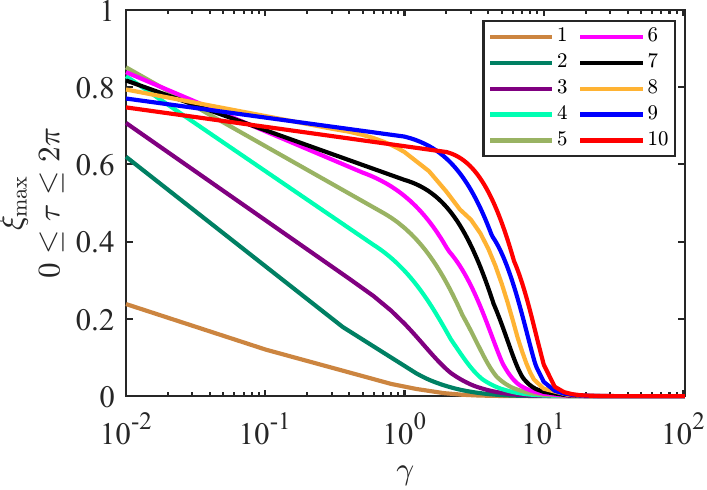} 
     \end{tabular} 
 \caption{ (Color online). The maximum ergotropy (in units of $\omega_q$) of the battery is measured over the charging interval $0 \le \tau \le 2\pi$, plotted as a function of the dissipation rate $\gamma$ (logarithmic scale) for different initial pure excitations of the charger. At very low coupling ($\gamma \ll 1$), the maximum ergotropy is determined by the initial excitation of the charger, reflecting more extractable work from higher excitations. For all initial excitations, ergotropy decreases sharply as $\gamma$ increases, reflecting the detrimental effect of charge on energy storage performance. Higher initial Fock numbers provide greater initial ergotropy, but they all converge to zero ergotropy beyond dissipation $\gamma=10$.}
 \label{sup4}
\end{figure}

Fig. \ref{sup4} depicts how much useful work could still be extracted while isolating the battery, as measured by the maximum ergotropy for dissipative Fock chargers. The measured ergotropy ($\xi_\text{max}$) describes the rapid depletion of usable energy from the battery caused by optimized dissipative processes during charging. These insights, like a pure Fock state, do not enhance the overall performance of the battery or the preparation of a sustainable charger. However, the thermalized Fock chargers with environmental coupling are capable of overcoming these difficulties, as presented in Fig. 3 in the main part of the manuscript.

\bibliographystyle{apsrev4-2}
\bibliography{ref}